\documentclass[12pt]{amsart}

\usepackage{epsfig}

\begin{document}

\title[]{\bf Perturbative Evaluation of the Effective Action
	     for a Self-Interacting Conformal Field on a Manifold with Boundary}

\author[]{George Tsoupros \\
       {\em School of Physics}\\
       {\em The University of New South Wales}\\
       {\em NSW 2052}\\
       {\em Australia}
}
\subjclass{49Q99, 81T15, 81T18, 81T20}
\thanks{present e-mail address: gts@gscas.ac.cn}

\begin{abstract}

In a series of three projects a new technique which allows for higher-loop
renormalisation on a manifold with boundary has been developed and used in 
order to assess the effects of the boundary on the dynamical behaviour of 
the theory. Commencing with a conceptual approach to the theoretical 
underpinnings of the, underlying, spherical formulation of Euclidean Quantum 
Field Theory this overview presents an outline of the stated technique's 
conceptual development, mathematical formalism and physical significance.     

\end{abstract}

\maketitle

{\bf I. Introduction}\\

The investigation of the effects generated on the dynamical behaviour of quantised 
matter fields by the presence of a boundary in the background geometry is an issue 
of central importance in Euclidean Quantum Gravity. This issue arises naturally in 
the context of any evaluation of radiative corrections to a semi-classical tunnelling 
geometry and has been studied at one-loop level through use of heat kernel and 
functional techniques. These methods were subsequently extended in the presence of 
matter couplings. Despite their success, however, such techniques have limited 
significance past one-loop order. Not only are explicit calculations of higher-order 
radiative effects far more reliable for the qualitative assessment of the theory's 
dynamical behaviour under conformal rescalings of the metric but they are, in addition, 
explicitly indicative of boundary related effects on that behaviour. Such higher-order 
calculations necessarily rely on diagrammatic techniques on a manifold with boundary. 
Fundamental in such a calculational context is the evaluation of the contribution which 
the boundary of the manifold has to the relativistic propagator of the relevant quantised 
matter field coupled to the manifold's semi-classical background geometry. It would be 
instructive, in this respect, to initiate an approach to such a higher loop-order 
renormalisation on a manifold with boundary by outlining the considerations which 
eventuated in the "background field" method, that approach to metric quantisation which is 
predicated on a fixed geometrical background.

The analysis relevant to the background field method can most easily be exemplified in 
the case of a massless scalar field minimally coupled to the background geometry. In 
the case of flat Euclidean space - defined by the analytical extension 
which eventuates in the replacement of $ x_0$ by $-ix_0$ - the generating functional 
relevant to the massless scalar field $ \phi$ coupled to a classical source $ J$ is 

\begin{equation}
Z[J] = \int D[\phi]e^{- \int d^4x[\frac{1}{2}\phi \partial^2 \phi -J\phi]}
\end{equation}
which, upon Gaussian integration yields

\begin{equation}
Z[J] = [det(\partial^2)]^{- \frac{1}{2}}e^{\int d^4xd^4y[J(x) \Delta(x,y) J(y)]}
\end{equation}
as a result of which, the scalar propagator of momentum $ k$

$$
x ---------y
$$
is

\begin{equation}
\Delta(x,y) = \int\frac{d^4k}{(2\pi)^4}\frac{e^{ik.(x-y)}}{k^2}
\end{equation}

The generating functional in the presence of gravity with a minimal coupling between 
the massless scalar field and the background metric $ g_{\mu\nu}$, which for the 
sake of mathematical consistency in the context of the present formalism is taken to 
have a Euclidean signature, is

\begin{equation}
Z[J] = \int D[\phi]e^{- \int d^4x \big{[}\frac{1}{2}\phi 
[(\frac{1}{\sqrt{g}}\partial_{\mu}g^{\mu\nu} \sqrt{g}\partial_{\nu})g] \phi -J\phi\sqrt{g}\big{]}}
\end{equation}
where the Riemannian manifold has been assumed, without loss of generality, to have no boundary so as 
to allow for vanishing surface terms. Again, Gaussian integration results in

\begin{equation}
Z[J] = [det(\frac{1}{\sqrt{g}}\partial_{\mu}g^{\mu\nu} \sqrt{g}\partial_{\nu})g]^{-\frac{1}{2}}
e^{\frac{1}{2}\int d^4xd^4y\big{[}J(x)G(x,y)J(y)\big{]}}
\end{equation}
which reveals the scalar propagator 

\begin{equation}
G(x,y) = <x|\big{[}(\frac{1}{\sqrt{g}}\partial_{\mu}g^{\mu\nu} \sqrt{g}\partial_{\nu})g]^{-1}|y>
\end{equation}
with $ |x>, |y>$ being orthonormal vectors in a suitably defined Hilbert space.

It would appear that the presence of such a propagator on the fixed geometrical background 
$ g_{\mu\nu}$ is consistent and would, for that matter, allow for the evaluation of scalar 
vacuum effects on condition of a tractable mathematical expression for the inverse to the 
associated metric-dependent kernel. Such an appearance, however, is physically irrelevant.
If the coupling between the scalar field and the background geometry is strong enough for 
the renormalisation group behaviour of the theory to justify the second quantisation of 
the former then, inevitably, the non-linear character of gravity in the context of the 
equivalence principle will render gravity, itself, just as much subject to second quantisation 
with the same degree of physical necessity \cite{Duff}. The quantisation of gravity enters, 
for that matter, non-trivially at all scales, a fact which necessitates a consistent
approach at least at the level of the quantisation of the background metric $ g_{\mu\nu}$. 
In the absence of a consistent quantum theory of gravity it would appear reasonable, in this 
respect, to pursue a linearised approach to second quantisation by quantising linear 
local perturbations $ h_{\mu\nu}$ of the background metric $ g^{c}_{\mu\nu}$ successfully
implemented through

\begin{equation}
g_{\mu\nu} = g^{c}_{\mu\nu} + h_{\mu\nu}
\end{equation}
so as to allow for $ h_{\mu\nu}$ to be treated as a null fluid in the stress tensor \cite{De Witt}.   

In the spirit of the background field approach to metric quantisation as outlined above the 
quantised linear perturbations $ h_{\mu\nu}$ represent the graviton, the quantum of the 
metric field propagating on the fixed background $ g^{c}_{\mu\nu}$. Naturally, in such an 
approach the weakness of the gravitational coupling $ G$ with respect to any other 
matter-field couplings at length scales well above the Planck scale allows for a consistent 
perturbative expansion with respect to the latter while keeping the former at the 
second-quantised zero order. The evaluation of such a perturbative expansion at high orders 
is predicated on a concrete mathematical expression for the matter-field propagator on       
$ g^{c}_{\mu\nu}$. Such an issue, is in general, non-trivial. Specifically, the inverse to 
the kernel for the scalar propagator in (6) does not readily admit a closed expression
in a general space-time. The only consistent approach in the content of the background field 
method is its treatment as a perturbative expansion through (7). The relevant diagrammatic 
representation involves both graviton-contributions which the scalar propagator on the 
background geometry of $ g^{c}_{\mu\nu}$ receives to all orders  

\begin{figure}[h]
\centering\epsfig{figure=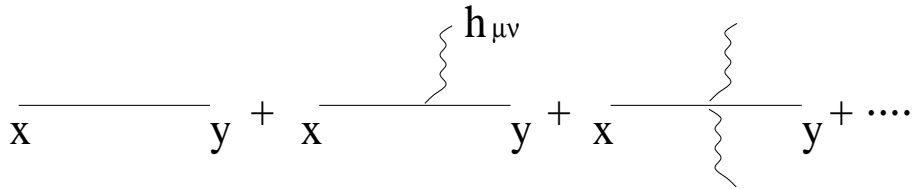, height = 25mm,width=120mm}
\caption{Graviton contributions to the scalar propagator}
\end{figure}
and scalar-contributions which the graviton propagator on the same background receives to all orders. 

The other dynamical aspect which (5) involves, in addition to that of the two 
propagators, stems from the associated determinant. Contrary to the 
Minkowski-space case expressed in (2), the 
metric-dependent determinant in (5) contributes non-trivially to the 
vacuum-to-vacuum 
amplitude expressed by $ Z[J]$. It has to be subjected, for that matter, 
to the same 
perturbative expansion through (7), an operation which is consistently 
accomplished through the exponentiation

\begin{equation}
e^{-\frac{1}{2}Trln(\frac{1}{\sqrt{g}}\partial_{\mu}g^{\mu\nu} \sqrt{g}\partial_{\nu})g}
\end{equation}

The trace in the exponentiated determinant results effectively 
in $ x \rightarrow y$
and generates, for that matter, a new diagrammatic representation with 
each of the previous 
scalar propagators closing upon itself  

\begin{figure}[h]
\centering\epsfig{figure=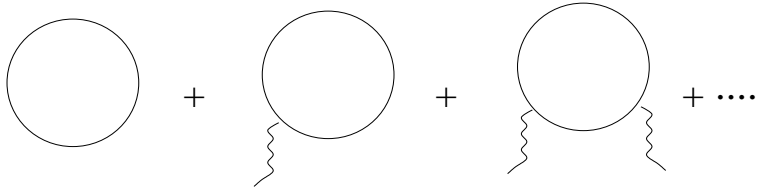, height = 25mm,width=100mm}
\caption{One-loop contributions to the effective action}
\end{figure}
as well as with each of the graviton propagators undergoing the 
same process.

The diagrammatic expansions for the scalar and graviton propagators 
as well as for that of the one-loop effective action correspond to   
representations derived through adiabatic expansions in Riemann 
normal coordinates and heat kernel techniques \cite{Birrell}. It 
is evident that, the necessary for perturbation, closed expression for 
such expansions is unattainable unless the relevant space-time is 
characterised by a high degree of symmetry. The natural candidate for 
such a space-time is the only maximally symmetric curved manifold,
the de Sitter space. The $ n$-dimensional de Sitter space can be 
represented as a hyperboloid 

\begin{equation}
z_0^2 - \sum_{i=1}^nz_ i^2 = -a^2
\end{equation}
embedded in a $ n+1$-dimensional Minkowski space with metric

\begin{equation}
ds^2 = dz_0^2 - \sum_{i=1}^ndz_ i^2 
\end{equation}

The Euclidean analogue of this space is a $ n$-dimensional  
sphere $ S_n$ of radius $ a$

\begin{equation}
\sum_{i=1}^{n+1}z_ i^2 = a^2
\end{equation}
embedded in a $ n+1$-dimensional Euclidean space. Since de Sitter
space is conformally equivalent to Minkowski space the action for 
a massless scalar field conformally coupled to the background 
geometry of $ S_n$ can readily be obtained by a conformal 
transformation of the corresponding exponentiated action featured 
in (1), in flat Euclidean space \cite{Drummond} 

\begin{equation}
S[\Phi] = \int d^n\sigma\big[\frac{1}{2}\frac{1}{2a^2}\Phi(L^2
- \frac{n(n-2)}{2})\Phi - \frac{\lambda}{\Gamma(p+1)}\Phi^p\big]
\end{equation}
with $ d^n\sigma=a^n\Omega_{n+1}$ being the volume element of 
the embedded $ S_n$, with $ p = \frac{2n}{n - 2}$ and with

\begin{equation}
L_{ab} = \eta_a\frac{\partial}{\partial \eta_b} - 
\eta_b\frac{\partial}{\partial \eta_a} 
\end{equation}
being the generator of rotations on $ S_n$ defined in terms of the embedding
vector $ \eta$. The formal equivalence between the spherically formulated 
scalar action obtained through the stated conformal transformation and the 
conformal scalar action defined directly on $ S_n$ has been shown 
\cite{McKeon Tsoupros}.  

It is evident that the closed expression which the kernel associated 
with the quadratic expression for the scalar field in (12) admits results, 
upon Gaussian integration over its exponentiated expression which replaces
that in (4), in an exact expression for the scalar propagator 
$ D(\eta, \eta')$ in (Euclidean) de Sitter space. Such an 
expression represents, in effect, an exact result for the formal 
summation of the graviton contributions to the scalar propagator

\begin{figure}[h]
\centering\epsfig{figure=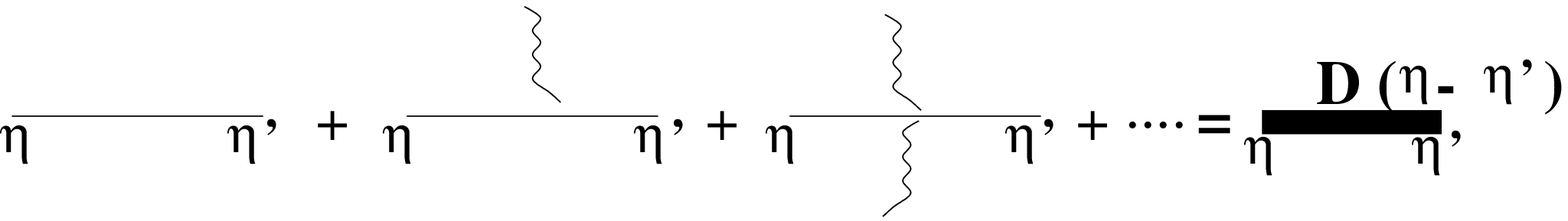, height = 25mm,width=160mm}
\caption{Exact propagator as a result of maximal symmetry}
\end{figure}
and is given by the elementary Haddamard function for propagation
between the space-time points $ \eta$ and $ \eta'$ on $ S_4$
\cite{Drummond}

\begin{equation}
D(\eta, \eta') = - \frac{1}{4{\pi}^2}\frac{1}{|\eta - \eta'|^2}
\end{equation}
The preceding analysis, effectively, reveals how maximal symmetry
renders the evaluation of higher-loop order vacuum effects and 
renormalisation tractable on a curved manifold. For example, the 
infinite series of diagrams in the case of a three-loop vacuum diagram
effected by the presence of a $ \lambda \phi^4$ self-interaction in a 
general space-time amounts on $ S_4$ to an exact expression for a 
three-loop diagram derived from the scalar propagator $ D(\eta, \eta')$ 

\begin{figure}[h]
\centering\epsfig{figure=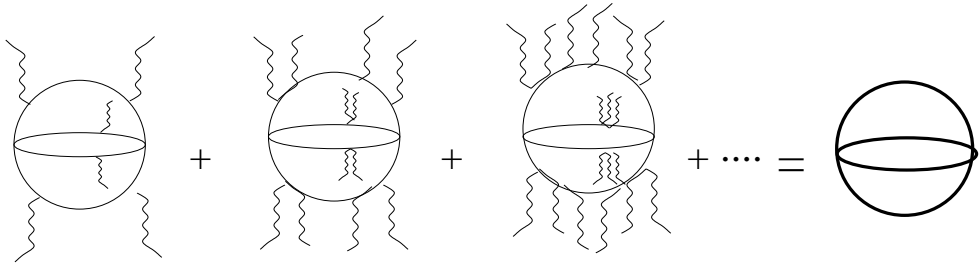, height = 35mm,width=130mm}
\caption{Higher-loop evaluation attained by maximal symmetry}
\end{figure}

In outlining the merit which maximal symmetry has for higher-loop 
renormalisation the preceding analysis calls into question the 
possibility of any extension of the concommitant spherical 
formulation to the physically important case of the spherical 
cap $ C_4$. The $ C_n$ is the n-dimensional Riemannian manifold 
of constant positive curvature embedded in an $ n+1$-dimensional 
Euclidean space and bounded by a $ n-1$-dimensional sphere of 
positive extrinsic curvature $ K$ (diverging normals). From the 
outset, such a geometric context is impervious to a direct 
application of the spherical formulation of a quantum field theory 
in four dimensions and, yet, suggestive of it. Specifically, the 
presence of a boundary in the manifold of constant positive curvature 
dispenses altogether with maximal symmetry. The remaining symmetry falls 
short of meeting the demand of expunging the theory of the mathematical 
complications hitherto discussed. It would be instructive, in this 
respect, to outline the fundamental differences which the direct 
application of the spherical formulation entails in the case of a
massless conformal field coupled to the background geometry of  
$ S_n$ and to that of $ C_n$ with a Dirichlet condition of a
vanishing value $ \Phi_{\partial C}=0$ on the boundary. The 
unbounded Laplace operator $ M$ defined on $ S_n$ being the 
kernel in the quadratic expression for the spherical action in 
(12)

\begin{equation}
M = \frac{1}{2}\frac{1}{2a^2}\big{[}
L^2 - \frac{n(n-2)}{2})\Phi - \frac{\lambda}{\Gamma(p+1)} 
\big{]}
\end{equation}
is conformally related to the corresponding d'Alambertian 
$ \partial^2$ in flat space-time and admits a complete
set of eigenfunctions 

\begin{equation}
MY_{\alpha}^N(\eta) = \lambda_N Y_{\alpha}^N(\eta)
\end{equation}
The latter are the spherical harmonics $ Y_{\alpha}^N(\eta)$ 
on $ S_n$ \cite{Drummond}. They are characterised by an integer 
degree $ N$ which is physically associated with the angular 
momentum flowing through the relevant propagator. In effect, 
the Green function $ D(\eta, \eta')$ associated with $ M$ 

\begin{equation}
MD(\eta, \eta') = \delta^{(n)}(\eta, \eta')
\end{equation}
admits an expansion in terms of $ Y_{\alpha}^N(\eta)$

\begin{equation}
D(\eta, \eta') = \sum_{N=0}^{\infty}\sum_{\alpha=0}^{N}
\frac{1}{\lambda_N}Y_{\alpha}^N(\eta)Y_{\alpha}^N(\eta')
\end{equation}    
which renders the spherical formulation a concrete mathematical 
content in configuration space primarily through the - fundamental
to perturbative calculations - formula

\begin{equation}
[(\eta - \eta')^2]^{\nu} = \sum_{N=0}^{\infty}\sum_{\alpha=0}^{N}
\frac{(2a)^{2\nu+n}{\pi}^{\frac{n}{2}}\Gamma(\nu+\frac{n}{2})\Gamma(N-\nu)}{
\Gamma(N+n+\nu)\Gamma(-\nu)}Y_{\alpha}^N({\eta})Y_{\alpha}^N({\eta}')
\end{equation}
In light of these mathematical underpinnings relevant to the 
spherical formulation in the stated case on $ S_n$ it becomes 
evident that any attempt for a direct application of the same 
formulation on $ C_n$ fails. Specifically, the action for the 
conformal scalar field although mathematically identical to (12)
generates an additional boundary term in the Einstein-Hilbert 
action $ S_{EH}$ involving the extrinsic curvature and the metric 
$ h_{ij}$ induced on the boundary with the stated Dirichlet 
condition. In effect, the action $ S$ for the theory on $ C_n$ is 
the additive result of (12) and the gravitational action 

\begin{equation}
S = S[\Phi] + S_{EH}
\end{equation}
with $ S_{EH}$ being for $ n=4$ \cite{George} 

\begin{equation}
S_{EH} = -\frac{1}{16\pi}\frac{1}{G}
\int_C d^4\sigma (R - 2\Lambda) + \int_{\partial C}
d^3x \sqrt{h}K\Phi^2
\end{equation}

As stated, this action is invariant under the set of conformal
rescalings of the metric

\begin{equation}
g_{\mu\nu} \rightarrow \Omega^2(x) g_{\mu\nu};~~~
\Phi \rightarrow \Omega^{1-\frac{n}{2}}\Phi
\end{equation}
This mathematical context reveals that, although the mathematical 
expression of the bounded spherical Laplace operator $ M_c$ on $ C_n$
remains the same as that of $ M$, its different domain generated by 
the presence of the boundary alters non-trivially the spectrum of 
eigenvalues thereby generating non-integer degrees $ N$. In effect, 
the $ n$-dimensional spherical harmonics $ Y_{\alpha}^N(\eta)$, 
although still eigenfunctions of $ M_c$, no longer form a complete set.
For that matter, configuration-space computations on $ C_n$ are 
substantially intractable as a result of the absence of the pivotal 
relation (19) \cite{George}. Moreover, the emergence of fractional 
degrees would tend to obscure the physical interpretation of any 
results attained by perturbative calculations. The presence of a 
boundary is, on the evidence, incompatible with a direct application of 
the spherical formulation.

It is evident, as a result of the preceding analysis, that the 
mathematical complications stemming from the presence of the boundary
relate directly to the associated eigenvalue problem. It would be 
desirable, for that matter, to relate the eigenvalue problem for the 
bounded Laplace operator $ M_c$ defined on $ C_n$ to that for the 
unbounded Laplace operator $ M$ on $ S_n$, the covering manifold of 
$ C_n$. The method of images is the simplest expedient to this end
and is predicated on the premise that the stated boundary effects  
which the propagator on $ S_n$ receives due to $ \partial C$ 
can, themeselves, be treated as equivalent to propagation on $ S_n$ in a 
manner which reproduces the boundary condition. In the present case of 
$ \Phi_{\partial C}=0$ the method of images results in \cite{George} 

\begin{equation}
D_{c}^{(n)}(\eta,{\eta}') = \frac{\Gamma(\frac{n}{2}-1)}{4\pi^{\frac{n}{2}}}
\big{[}\frac{1}{|{\eta}-{\eta}'|^{n-2}} -
\frac{1}{|\frac{a_{{\eta}'}}{a_B}{\eta}-
\frac{a_B}{a_{{\eta}'}}{\eta}'|^{n-2}}\big{]}
\end{equation}
for the conformal scalar propagator with $ a_{{\eta}'}$ being the 
geodesic distance between the cap's pole and space-time point $ {\eta}'$
The merit of the method of images is explicit. The 
fundamental part of the propagator, being identical to (14), signifies 
propagation on $ S_n$ whereas the boundary part, which expresses the 
contributions of $ \partial C$ on the fundamental part, can be seen 
to signify, itself, propagation on $ S_n$ between the associated 
space-time points $ \frac{a_{{\eta}'}}{a_B}{\eta}$ and $
\frac{a_B}{a_{{\eta}'}}{\eta}'$. In effect, although as stated 
$ D_{c}^{(n)}(\eta,{\eta}')$ is impervious to an expansion on the
lines of (18) both its fundamental and boundary components admit,
in principle, such an expansion. Herein the merit of this technique
lies. The spherical formulation emerges independently for both
components of the propagator on $ C_n$. Specifically, the image 
propagator on $ S_n$ associated with the boundary component of 
$ D_{c}^{(n)}(\eta,{\eta}')$ also admits a desired expansion on the 
lines of (19). However, although the limit of vanishing geodesic 
separations at the coincidence limit $ \eta \rightarrow \eta'$ is, 
as expected, inherent in the singular component the same limit is
unattainable by the boundary component which remains, effectively, 
always finite - as was mathematically expected from the boundary 
part of the Green function $ D_{c}^{(n)}(\eta,{\eta}')$. That, in turn, 
enforces upon the stated expansion for image propagation the 
condition of vanishing propagation for geodesic distances smaller 
than $ |\frac{a_{{\eta}'}}{a_B}{\eta}'-\frac{a_B}{a_{{\eta}'}}{\eta}'|$. 
In effect, the - equivalent to (19) - expansion for image propagation is 
\cite{Tsoupros}

\begin{equation}
[|\frac{a_{{\eta}'}}{a_B}{\eta}- \frac{a_B}{a_{{\eta}'}}{\eta}'|^2]^{\nu} = 
\sum_{N=0}^{N_0}\sum_{\alpha=0}^{N}
\frac{(2a)^{2\nu+n}{\pi}^{\frac{n}{2}}\Gamma(\nu+\frac{n}{2}+
\frac{1}{N_0})\Gamma(N-\nu+ \frac{1}{N_0})}{
\Gamma(N+n+\nu+ \frac{1}{N_0})\Gamma(-\nu)}Y_{\alpha}^N({\eta})Y_{\alpha}^N({\eta}')
\end{equation}  
where use has been made of 

\begin{equation}
\sum_{\alpha=0}^{N}Y_{\alpha}^N(\frac{a_{{\eta}'}}{a_B}{\eta})
Y_{\alpha}^N(\frac{a_B}{a_{{\eta}'}}{\eta}') = \sum_{\alpha=0}^{N}
Y_{\alpha}^N({\eta})Y_{\alpha}^N({\eta}') 
\end{equation}
The inverse $ \frac{1}{N_0}$ of the upper limit related to the cut-off 
angular momentum $ N_0$ for image propagation in the arguments of the 
$ \Gamma$ functions ensures the finite character of (24) at $ n \rightarrow 4$ 
and is expected on the grounds that the situation of scalar propagation towards 
the boundary - that is, that situation characterised by the limit 
$ a_{\eta'} \rightarrow a_B$ - results in $ N_0 \rightarrow \infty$ at the 
limit of vanishing geodesic separations $ \eta \rightarrow \eta'$ in which 
case, the expansion in (24) is reduced to the exact expansion in (19). 

The perturbative evaluations of radiative effects in the configuration space of 
$ C_4$ within the mathematical context hitherto outlined necessitate the further 
specification of the regulating scheme for the concommitant divergences as 
well as a concrete expression for the necessary integration over the relevant 
vertices. The definition of the theory in $ n$ dimensions is, in fact, suggestive 
of the technique of dimensional regularisation. The latter manifests all 
divergences arising from the Feynman integrals at the limit 
$ \eta \rightarrow \eta'$ as poles at the dimensional limit of 
$ n \rightarrow 4$ after an analytical extension of the space-time dimensionality
$ n$. In configuration space any diagramatic calculation on $ C_n$ involves powers 
of the propagator $ D_{c}^{(n)}(\eta,{\eta}')$ and, for that matter, products 
featuring powers of its fundamental part $ |{\eta}-{\eta}'|^{2-n}$ and of its 
boundary part 
$ |\frac{a_{{\eta}'}}{a_B}{\eta} - \frac{a_B}{a_{{\eta}'}}{\eta}'|^{2-n}$. The 
former are evaluated through use of (19) and the latter through use of (24). An
immediate consequence of (24) is the definite finite character of any power of the 
boundary part of the propagator. In effect, in the context of dimensional 
regularisation, all possible divergences at the dimensional limit 
$ n \rightarrow 4$ stem exclusively from the fundamental-part related expansion 
in (19). This analysis reveals the mathematical origin and physical character of the 
pole structures in configuration space. These structures arise within the 
context of integrations of a diagram's vertices over the relevant manifold's volume 
according to the Feynman rules. It can readily be seen from (19) and (24) that such 
integrations invariably eventuate in the expression 
$ \int_Cd^{n}{\eta}Y_{\alpha}^N(\eta)Y_{\alpha'}^{N'}(\eta)$. The smaller symmetry 
associated with the geometry of $ C_n$ precludes the orthonormality condition
which emerges for that integral on $ S_n$. The result, attained through two successive
applications of the divergence theorem, is instead \cite{Tsoupros}

\begin{equation}
\int_C d^{n}{\eta}Y_{\alpha}^N(\eta)Y_{\alpha'}^{N'}(\eta) = Aa^2
\oint_{\partial C} d^{n-1}{\eta}[KY_{\alpha}^N(\eta)Y_{\alpha'}^{N'}(\eta) + 
2n_pY_{\alpha'}^{N'}(\eta)D_pY_{\alpha}^{N}(\eta)]
\end{equation}
with 

\begin{equation}
A = \frac{1}{(N'+\frac{n}{2}-1)(N'+\frac{n}{2}) - (N+\frac{n}{2}-1)(N+\frac{n}{2})};~~N \neq N' 
\end{equation}
and with each surface integral admitting a concrete, although involved, expression in terms of
Gegenbauer polynomials. In addition to the radius $ a$ of the embedded $ \it{C_n}$ this 
expression features, as expected, the extrinsic curvature of $ \partial C_n$. Evidently, it is 
exactly this feature of any diagrammatic calculation in configuration space which, as will be
explicitly shown, ensures a simultaneous renormalisation of boundary and surface terms in the 
effective action at any specific loop order. The immediate issue which such simultaneity calls
into question is the perturbative generation of surface counterterms in addition to possible 
novel volume counterterms by vacuum scalar processes in the gravitational effective action on 
$ C_4$. As the exclusive source of gravitational counterterms is the zero-point function, an 
outline of the main aspects of its perturbative evaluation to $ O(\hbar^3)$ would illustrate 
the highly non-trivial effects of the presence of a boundary on the theory as well as provide 
an essential demonstration of the merit of the techniques hitherto analysed \cite{GT}.

On a general, unbounded, four dimensional manifold the bare gravitational action assumes the
form \cite{Birrell}

\begin{equation}
S_g = \int d^4x\sqrt{-g}\big{[}\Lambda_0 + \kappa_0 R + a_0 G + b_0 H + c_0 R^2 \big{]}  
\end{equation}
with  

\begin{equation}
G=R_{abcd}R^{abcd}-4R_{ab}R^{ab}+R^2~~;~~H=C_{abcd}C^{abcd}
\end{equation}
Additional terms are expected in the presence of a boundary. 

The first two terms in the perturbative expansion of the zero-point function in powers of 
$ \hbar$ (loop expansion) are diagramatically represented by the graphs of fig.(5).

\begin{figure}[h]
\centering\epsfig{figure=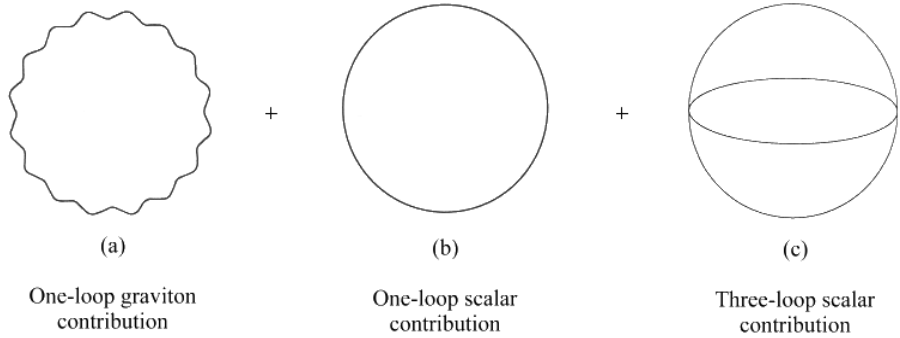, height = 45mm,width=120mm}
\caption{$ O(\hbar^3)$-related contributions to the zero-point function}
\end{figure}
The ``bubble'' diagrams in fig.(5a) and fig.(5b) account for the one-loop contribution
to the zero-point function of the theory. Their simultaneous presence in any curved 
space-time is expected on the basis of general theoretical considerations \cite{Birrell}, \cite{Duff}. 
They are characterised by the absence of interaction vertices and, on power counting grounds, 
are responsible for the simultaneous one-loop contributions to volume and boundary 
effective Einstein-Hilbert action on any manifold with boundary. They have been shown to be 
finite provided that dimensional regularisation is used \cite{George}. In effect, to 
$ O(\hbar^3)$ the exclusive source of contributions to the bare cosmological constant 
$ \Lambda_0$ and the bare gravitational couplings $ \kappa_0, a_0, b_0, c_0$ in (28) on 
a general manifold is the diagram in fig.(5c) representing the term 

\begin{equation}
I_c = \lambda^2 \int d^n{\eta}d^n{\eta'} [D_{c}^{(n)}(\eta,{\eta}') ]^4
\end{equation}
On $ C_4$, the Euclidean de Sitter space, the relation \cite{McKeon Tsoupros}

\begin{equation}
R = \frac{n(n -1)}{a^2}
\end{equation}
effectively reduces (28) to 

\begin{equation}
S_g = \int d\sigma \big{[}\Lambda_0 + \kappa_0 R + c_0 R^2 \big{]}  
\end{equation} 
In such a context the stated issue of generation of additional counterterms in $ S_g$ translates 
to the possibility of additional counterterms generated by the diagram of fig.(5c) on $ C_4$. A direct 
replacement of (23) into (30) resolves the double integral with respect to the vertices of the 
diagram into five such integrals over products of terms of the form (19) and (24). The evaluation 
of these integrals in the context of (26) and its associated expressions results, at 
$ n\rightarrow 4$ $ (\epsilon \rightarrow 0)$ in a substantially involved expression for 
$ I_c$ \cite{GT} \footnote{For the sake of conformity with length limitations, only an outline of 
this expression's physical significance is cited.}

\newpage

$$
I_c = 
\lambda^2\frac{1}{\pi^{4}}\frac{1}{2^8}\sum_{N=0}^{\infty}\big{[}
\frac{1}{\pi^{2}}\frac{1}{3^3}\frac{1}{2^7}\frac{(N+2)(N+3)}{(N+1)(N+4)} 
\big{(}C_{N+1}^{\frac{3}{2}}(cos{\theta_4^0})\big{)}^2 + $$

$$
\sum_{N'=0}^{N'_0}\frac{1}{[{N'}^2-N^2+3(N'-N)]^2}[\Gamma(\frac{1}{N'_0}) -
\frac{1}{3}\frac{1}{2}\frac{(\Gamma(1+\frac{1}{N'_0})\Gamma(N'+1+\frac{1}{N'_0})}{\Gamma(N'+3+\frac{1}{N'_0})}(N+1)(N+2)] \times $$

$$
3\frac{1}{2^3}\frac{1}{\pi^2}
F^2N'\big{]}\frac{RK^2}{\epsilon}\int_Cd^4{\eta} + $$

$$
\lambda^2\frac{1}{\pi^{6}}\frac{1}{2^{13}}\sum_{N=0}^{\infty}\sum_{N'=0}^{N'_0}
\frac{1}{[{N'}^2-N^2+3(N'-N)]^2} \times $$

$$
\left[ \Gamma(\frac{1}{N'_0}) - 
\frac{1}{3}\frac{1}{2}\frac{(\Gamma(1+\frac{1}{N'_0})\Gamma(N'+1+\frac{1}{N'_0})}{\Gamma(N'+3+\frac{1}{N'_0})}(N+1)(N+2) \right](BH)^2
\frac{R^2}{\epsilon}\int_Cd^4{\eta} +  $$

$$
\lambda^2\frac{1}{\pi^{6}}\frac{1}{2^{8}}\sum_{N=0}^{\infty}\sum_{N'=0}^{N'_0}
\frac{1}{[{N'}^2-N^2+3(N'-N)]^2} \times $$

$$
\left[ \Gamma(\frac{1}{N'_0}) - 
\frac{1}{3}\frac{1}{2}\frac{(\Gamma(1+\frac{1}{N'_0})\Gamma(N'+1+\frac{1}{N'_0})}{\Gamma(N'+3+\frac{1}{N'_0})}(N+1)(N+2) \right]\times $$

\begin{equation}
(FBH)N'\big{(}sin(\theta_4^0)\big{)}^{-3}
\frac{RK}{\epsilon}\oint_{\partial C}d^3{\eta} 
\end{equation}
The three sectors explicitly featured in this expression are indicative of the 
theory's dynamical behaviour on $ C_4$. The qualitatively new features which
the presence of the boundary engenders break the classical conformal invariance 
and dissociate completely that dynamical behaviour from its corresponding one on 
$ S_4$ primarily through the generation of topology-related divergences \cite{GT}. 
There are two volume terms proportional to $ RK^2$ and $ R^2$ respectively as 
well as a surface term proportional to $ RK$. The $ R^2$ term yields a direct 
contribution to $ c_0$ in the corresponding sector of (32). However, the $ RK^2$ 
term signifies a qualitatively new quantum correction in the bare and, for that 
matter, effective gravitational action. As $ K^2$ indicates it is non-trivially 
generated by $ \partial C$. The corresponding $ RK$-related sector in the bare 
gravitational action signifies, indeed, the first surface counterterm generated 
by vacuum effects on a general manifold with boundary. Moreover, the simultaneous 
emergence of boundary and surface divergences confirms the theory's potential for 
a simoultaneous renormalisation of boundary and surface terms in higher loop-orders. 

{\bf Acknowledgement}

I wish to acknowledge the discrimination, confrontation and contempt with which
the ``lucky country'' received my research effort all these years. My contribution 
to theoretical research would have never had its present significance without 
Australia's mindless rejection.

\end{document}